\documentclass[fleqn,usenatbib]{mnras}

\usepackage{newtxtext,newtxmath}

\usepackage[T1]{fontenc}

\usepackage{graphicx}	
\usepackage{amsmath}	
\usepackage{soul}
\usepackage{xcolor}

\title[Carbon-oxygen ultra-massive white dwarfs in general relativity]{Carbon-oxygen ultra-massive white dwarfs in general relativity \thanks{The cooling sequences are publicly available at \href{ http://evolgroup.fcaglp.unlp.edu.ar/TRACKS/tracks.html}{http://evolgroup.fcaglp.unlp.edu.ar/TRACKS/tracks.html}}}

\author[Althaus et al.]{
Leandro G. Althaus$^{1,2}$\thanks{E-mail: althaus@fcaglp.unlp.edu.ar},
Alejandro H. C\'orsico$^{1,2}$,
Mar\'ia E. Camisassa$^{3}$,
Santiago Torres$^{3,4}$,
\newauthor 
Pilar Gil-Pons$^{3,4}$, 
Alberto Rebassa-Mansergas$^{3,4}$ and
Roberto Raddi$^{3}$
\\
$^{1}$  Facultad de Ciencias Astron\'omicas y Geof\'{\i}sicas, 
           Universidad Nacional de La Plata, 
           Paseo del Bosque s/n, 1900 
           La Plata, 
           Argentina\\
$^{2}$ Instituto de Astrof\'isica La Plata, UNLP-CONICET,
           Paseo del Bosque s/n, 
           1900 La Plata, 
           Argentina\\
$^{3}$ Departament de F\'\i sica, 
           Universitat Polit\`ecnica de Catalunya, 
           c/Esteve Terrades 5, 
           08860 Castelldefels, 
           Spain\\
$^{4}$   Institute for Space Studies of Catalonia, 
           c/Gran Capita 2--4, 
           Edif. Nexus 104, 
           08034 Barcelona, 
           Spain
}         
\date{Accepted XXX. Received YYY; in original form ZZZ}

\pubyear{2023}

\begin{document}
\label{firstpage}
\pagerange{\pageref{firstpage}--\pageref{lastpage}}
\maketitle

\begin{abstract}
We employ the La  Plata  stellar  evolution  code, {\tt  LPCODE}, to compute the
first set of constant rest-mass carbon-oxygen ultra-massive white dwarf evolutionary
sequences for masses higher than 1.29\ M$_{\sun}$  that
fully take  into account  the effects of  general relativity  on their
structural   and  evolutionary   properties. In  addition, we  employ the  {\tt
  LP-PUL}  pulsation  code  to compute  adiabatic  $g$-mode  Newtonian
pulsations on  our fully relativistic equilibrium  white dwarf models.
We  find  that carbon-oxygen  white  dwarfs  more massive  than  1.382\ 
M$_{\sun}$ become  gravitationally unstable  with respect  to general
relativity effects, being this limit higher than the 1.369\  M$_{\sun}$
we found for oxygen-neon white  dwarfs.  As the stellar mass approaches
the limiting mass  value, the stellar radius becomes substantially smaller compared with the Newtonian models.  
Also, the thermo-mechanical  and evolutionary  properties  of  the most  massive
white dwarfs are strongly affected  by general relativity effects.  We
also  provide magnitudes  for our  cooling
sequences in  different passbands. Finally,  we explore for  the first
time the  pulsational properties  of relativistic  ultra-massive white
dwarfs  and find  that  the period  spacings  and oscillation  kinetic
energies  are strongly  affected in  the  case of  most massive  white
dwarfs.  We  conclude that  the general  relativity effects  should be
taken  into account  for  an accurate  assessment  of the  structural,
evolutionary, and  pulsational properties of white  dwarfs with masses
above $\sim$  1.30\ M$_{\sun}$.   
\end{abstract}



\section{Introduction}
\label{introduction}

Ultra-massive white dwarf (UMWD) stars,  defined as those white dwarfs with  masses higher than $\simeq  1.05$   M$_{\sun}$,  are   involved  in   a  variety   of  extreme
astrophysical  phenomena  such  as   type  Ia  supernovae,  micronovae
explosions,  radio   transients  via  an   accretion-induced  collapse
of the white dwarf \citep{2019MNRAS.490.1166M},  and the  formation  of
millisecond pulsars  \citep {2022MNRAS.510.6011W}, as well  as stellar
mergers \citep{2020ApJ...891..160C}. 
They are also key objects 
for the study of  the theory of high-density plasmas, general
relativity,    and   to    constrain    the    emission   of    axions
\citep{2022PhRvL.128g1102D}. Among the UMWDs, those in the highest mass range 
($\gtrsim 1.30$ M$_{\sun}$) are crucial to constrain the initial mass threshold
for the formation of white dwarfs and the electron-capture or core-collapse 
supernovae \citep{doherty2017}. 

The  existence of  UMWDs  has  been reported  in
several                                                        studies
\citep{2004ApJ...607..982M,2016IAUFM..29B.493N,2011ApJ...743..138G,2013ApJS..204....5K,2015MNRAS.450.3966B,2016MNRAS.455.3413K,2017MNRAS.468..239C,2021MNRAS.503.5397K,Hollands2020,2021Natur.595...39C,Jimenez2023}. In
particular,  the  number  of  UMWDs  with  mass
determinations  beyond   about  1.30  M$_{\sun}$   has  substantially
increased  in recent  years. For  instance, \cite{2018ApJ...861L..13G}
derived a  mass of  $1.28\pm0.08$ M$_{\sun}$ for the  long-known white
dwarf GD 50.  \cite{2020MNRAS.499L..21P} discovered a rapidly rotating
UMWD,    WDJ183202.83+085636.24,     with
$M=1.33\pm0.01$ M$_{\sun;}$    meanwhile,   \cite{2021Natur.595...39C}
reported  the  existence  of  a highly  magnetised,  rapidly  rotating UMWD,  ZTF  J190132.9+145808.7, with  a mass  of
$\sim 1.327 - 1.365$  M$_{\sun}$.  \cite{2021MNRAS.503.5397K} concluded
that other 22  white dwarfs in the solar neighbourhood  could also have
masses over  1.29 M$_{\sun}$,  if they  had pure  H envelopes  and CO
cores.

White dwarfs with H-rich atmospheres, that is DA spectral type, and effective temperatures in the range $10\,500 \lesssim T_{  \text{eff}} \lesssim 13\,000$ K exhibit pulsations due to spheroidal non-radial gravity($g$) modes \citep{2008ARA&A..46..157W, 2008PASP..120.1043F, 2010A&ARv..18..471A, 2019A&ARv..27....7C}. These pulsating stars are called ZZ Ceti or DAV
white dwarfs and constitute the most populated class of pulsating white dwarfs. Although the vast majority of ZZ Ceti stars have stellar masses in the range 
$0.5 \lesssim M_{\star}/{\text{M}}_{\sun} \lesssim 0.8$, pulsations have also been detected in, at least, four ultra-massive ZZ Ceti stars. They are BPM~37093 \citep[$M_{\star}=   1.1\ {\text{M}}_{\sun}$,][]{1992ApJ...390L..89K},  GD~518  \citep[$M_{\star}=
  1.24\ {\text{M}}_{\sun}$,][]{2013ApJ...771L...2H}, SDSS~J084021
\citep[$M_{\star}= 1.16\ {\text{M}}_{\sun}$,][]{2017MNRAS.468..239C}, and
WD~J212402 \citep[$M_{\star}= 1.16\ {\text{M}}_{\sun}$,][]{2019MNRAS.486.4574R}.
Recently, \cite{2023arXiv230410330K} have reported the discovery of 
the fifth known ultra-massive ZZ Ceti star, the DAV star WD~J004917.14$-$252556.81, which,
with $M_{\star}$ between  $\sim 1.25\ {\text{M}}_{\sun}$ (for an ONe core) and $\sim 1.30\ {\text{M}}_{\sun}$ 
(for a CO core) is  the most massive pulsating white dwarf currently known. 
It is likely that pulsating white dwarfs as massive as WD~J004917.14$-$252556.81 or even more massive will be identified in the coming years with the advent of huge volumes
of high-quality photometric data collected by space missions such as
the ongoing {\it TESS} mission \citep{2014SPIE.9143E..20R} and the
future {\it PLATO} space telescope \citep{2014ExA....38..249R}. 

The increasing number of
UMWDs with masses beyond $\sim 1.30\ {\text{M}}_{\sun}$, as well as the
immediate prospect of detecting pulsating white dwarfs with such masses, demand
new appropriate theoretical evolutionary models to analyse them. Such new models must revise existing progenitor formation scenarios and white dwarf evolution, and contemplate the introduction of more realistic input physics.
In particular, it is necessary to calculate models that take into account
general relativity effects and to evaluate the resulting pulsational properties.

Theoretical  evolutionary scenarios  predict  an  oxygen-neon (ONe)  or
carbon-oxygen (CO)  core-chemical composition for  UMWDs.   ONe  core white  dwarfs  are  expected  to result  from  the
semi-degenerate carbon burning during the single evolution of massive
intermediate-mass  stars that    evolve    to    the     super    asymptotic    giant    branch. The initial mass threshold for carbon burning is around 6–9\ M$_{\sun}$, and was reported to
depend mainly on  the metallicity and the treatment  of convective boundaries
\citep{1997ApJ...485..765G,2005A&A...433.1037G,2006A&A...448..717S,
2011MNRAS.410.2760V,doherty2015}. The effects of rotation and 
nuclear reaction rates are also known to affect this mass threshold \citep{doherty2017}. 
The formation of an ONe core composition in UMWDs is also
predicted by the merger of two  white dwarfs as a result of off-center
carbon burning in  the merged remnant when the remnant  mass is larger
than  1.05 M$_{\sun}$  \citep{2021ApJ...906...53S}.

On  the other  hand, CO  UMWDs  can result  from
single   stellar   evolution  \cite[see][]{2021A&A...646A..30A}.    In
particular,   these  authors  have shown  that   isolated  progenitors   of
UMWDs can avoid C burning on the asymptotic giant
branch,  supporting the  existence of  CO cores  in white  dwarfs more
massive than  1.05 M$_{\sun}$.   Specifically, reduced wind  rates with
respect to  the standard prescriptions  and/or the occurrence  of core
rotation   (which    lowers   the   internal   pressure,    see   also
\citealt{dominguez1996})   hamper   C-ignition,  thus   favouring   the
formation of UMWDs with CO cores. In addition, the formation of  UMWDs with  CO cores can also result from white dwarf  merger, see
\cite{2022MNRAS.512.2972W}. These author have shown that CO UMWDs with masses below  1.20 M$_{\sun}$ can be produced from 
massive CO white dwarfs + He white dwarf merger. The merger scenario for the formation of UMWDs is sustained by the fact that  a considerable fraction of  the massive white
dwarf  population  is thought to be formed   as  a   result  of   stellar  mergers
\citep{2020A&A...636A..31T,2020ApJ...891..160C,2022MNRAS.511.5462T}.

The study of the evolution of  UMWDs has been the
subject    of     numerous    recent    papers.      In    particular,
\cite{2019A&A...625A..87C}  studied in detail the  full evolution
of ONe UMWDs with masses up  to $1.29\, {\text{M}}_{\sun}$
considering realistic initial chemical profiles that are the result of
the      full       progenitor      evolution       calculated      in
\cite{2010A&A...512A..10S}.  More recently, \cite{2021A&A...646A..30A}
and \cite{2022MNRAS.511.5198C} computed the evolution of UMWDs up  to $1.29\, {\text{M}}_{\sun}$ with CO cores  resulting from the
complete  evolution   of  single  progenitor  stars   that  avoid  the
C-ignition    during    the     super    asymptotic    giant    branch
\cite[see][]{2021A&A...646A..30A}. Finally, \cite{2021ApJ...916..119S}
 studied  the evolution of  white dwarfs more massive  than $1.29\,
{\text{M}}_{\sun}$,  with the  focus on  neutrino cooling  via the  Urca process,
showing that  this process is  important for the age  determination of
ONe core white dwarf stars.

For  the  most  massive  white   dwarfs,  the  importance  of  general
relativity  for their  structure  and evolution  cannot be  completely
disregarded.   In fact,  numerous works  based on  static white  dwarf
structures have  shown that general relativistic  effects are relevant
for  the  determination   of  the  radius  of   massive  white  dwarfs
\citep{2011PhRvD..84h4007R,2017RAA....17...61M,2018GReGr..50...38C,
  2021ApJ...921..138N}. In particular,  these studies demonstrated that
for  fixed values  of    mass, deviations  up  to  $50\%$ in  the
Newtonian  white dwarf  radius are  expected compared  to the  general
relativistic  white  dwarf  radius.    More  recently,
\cite{2022A&A...668A..58A}   have presented  the
first  set  of  constant   rest-mass  ONe UMWD
evolutionary models more massive than 1.30 M$_{\sun}$
that fully take  into account the effects of  general relativity. This
study demonstrates that  the general relativity effects  must be taken
into account to  assess the structural and  evolutionary properties of
the most  massive white dwarfs.

In this paper, we extend the relativistic calculations presented in 
\cite{2022A&A...668A..58A} to the case of
  UMWDs with CO cores that result from the complete  evolution   of  single
progenitor  stars   that  avoid C-ignition \cite[see][]{2021A&A...646A..30A}.
We employ the La  Plata stellar evolution code, {\tt  LPCODE}, to compute
full  evolutionary sequences of  white dwarfs more massive than 1.29 M$_{\sun}$, taking into account general relativity effects.
For  comparison  purposes, additional sequences of identical initial models 
are computed for the Newtonian gravity  case. We also
employ the {\tt LP-PUL} pulsation code to perform an exploratory pulsational investigation by computing $g$-mode Newtonian
pulsations on fully relativistic equilibrium white dwarf models and comparing the results of the period spacings and oscillation kinetic energies with the case of Newtonian white dwarf models.
The strong effects of general relativity on the thermo-mechanical structure of the 
most massive CO white dwarfs we report in this work may impact the energetics of the explosion 
and the nucleosynthesis of neutron-rich matter, and thus be relevant 
for the better understanding of type-Ia supernova explosions. Furthermore, 
general relativity effects could affect the critical density that separates 
the explosive and the collapse outcomes of these supernovae \citep{1999MNRAS.307..984B}.

This  paper  is organised  as  follows.
In  Sect.  \ref{equations} we  provide the  main
computational   details  and   input  physics   of  our  relativistic
white  dwarf
sequences. Sect.  \ref{results} is  devoted to explore the impact  of general
relativity  effects on  the  relevant evolutionary  properties of  our
massive white dwarfs. In this section  we also compare  the
predictions of our  new white dwarf sequences  with observational data
of  UMWDs.  In Sect. \ref{pulsational} we assess the Newtonian pulsational properties of 
relativistic white dwarf models, and compare them with the pulsational properties of Newtonian models.  Finally,   in
Sect. \ref{conclusions} we summarise the main findings of this work.

\section{Computational details and input physics}
\label{equations}

The set  of CO UMWD evolutionary  sequences was
computed with the stellar evolution  code {\tt LPCODE}, developed by the La
Plata group \citep{2005A&A...435..631A, 2013A&A...555A..96S, 2015A&A...576A...9A,
2016A&A...588A..25M,2020A&A...635A.164S,
2020A&A...635A.165C}. It has been modified to include the effects of general
relativity,      following      the       formalism      given      in
\cite{1977ApJ...212..825T}, see  \cite{2022A&A...668A..58A} for details about the implementation
of pertinent general relativity equations in  {\tt LPCODE}. In particular,
the  fully general
relativistic partial differential equations governing the evolution of
a spherically symmetric star are formulated in a way that they resemble
the    standard    Newtonian    equations   of    stellar    structure
and evolution. Of relevance in this formulation are the dimensionless general
relativistic correction factors that are applied on these equations, that is $\mathscr{H}, \mathscr{G},  \mathscr{V}$, and $\mathscr{R}$,
which correspond to  the enthalpy,  gravitational  acceleration,  volume,
and  redshift correction factors, respectively. These factors turn to
unity in the Newtonian  limit, and are given by the expressions:

\begin{align}
\mathscr{H} &= \frac{\varrho^t}{\varrho} + \frac{P}{\varrho c^2},\\
\mathscr{G} &= \frac{ m^t + 4 \pi r^3 P/c^2} {m},   \\
\mathscr{V} &= \left(1 - \frac{ 2 G m^t}{ r c^2}\right)^{-1/2}, \label{eq_V} \\
\mathscr{R} &= e^{\Phi/c^2}, 
\label{R-fact}
\end{align}

\noindent 
where $m$ is the  rest mass inside a radius $r$ or
baryonic mass, $\varrho$
is the density of rest mass,  $m^t$ is the mass-energy inside $r$
and includes contributions from the rest-mass energy, the internal
energy,   and   the   gravitational   potential   energy,   which   is
negative. Since the internal and gravitational potential energy change
during  the course  of evolution,  the  stellar mass-energy  is not  a
conserved   quantity.   $\varrho^t$   is    the   density   of  the total
non-gravitational mass-energy,  and includes  the density of the rest mass
plus contributions  from the kinetic and  potential energy densities  due to
particle interactions (it does not include the gravitational potential
energy density),  that is, $\varrho^t=  \varrho + (u \varrho)/  c^2 $,
where $u$ is the internal energy per unit mass.
$\Phi$ is the general  relativistic gravitational potential related to
the temporal metric coefficient.

\begin{table*}
   \centering
  \caption{Relevant characteristics of our sequences at $T_{ \text{eff}}=10\,000\,$K.  $M_{\text{WD}}$: total baryonic
  mass.  $M_{\text{G}}$: total gravitational mass. $R^{\text{Newt}}$: stellar radius in the Newtonian case. $R^{\text{GR}}$: 
  stellar radius in the general relativity case. g$^{\text{Newt}}$: surface gravity in the Newtonian case. g$^{\text{GR}}$: surface
  gravity in the general relativity case. $\varrho_c^{\text{Newt}}$: central density  in the Newtonian case.  $\varrho_c^{\text{GR}}$:
  central density of rest mass in the general relativity case. }
  \label{table1}
\begin{tabular}{lccccccc} 
  \hline
$M_{\text{WD}}$ & $M_{\text{G}}$  & $R^{\text{Newt}}$ & $R^{\text{GR}}$  & log g$^{\text{Newt}}$ &  log g$^{\text{GR}}$ & $\varrho_c^{\text{Newt}}$  & $\varrho_c^{\text{GR}}$\\
  $M_{\sun}$ &   $M_{\sun}$ &  $R_{\sun}$ &  $R_{\sun}$  &  cm s$^{-2}$  &    cm s$^{-2}$   & g cm$^{-3}$   & g cm$^{-3}$  \\
  \hline
  1.29   & 1.28978 &  0.00422592  &  0.00412784  &  9.296     &  9.317  &  $ 4.88\times 10^{8}$  &    $5.34 \times 10^{8}$     \\
  1.31   & 1.30977 &  0.00389625 &   0.00377446 &   9.374    &   9.402  &   $ 6.74 \times 10^{8}$  &   $7.61 \times 10^{8}$     \\
  1.33   & 1.32975 &  0.00353773 &   0.00337825  &  9.464     &  9.505   &   $ 9.84 \times 10^{8}$  &   $1.17 \times 10^{9}$    \\
  1.35   & 1.34974 &  0.00313854 &   0.00291305  &  9.574    &   9.640  &   $ 1.56 \times 10^{9}$ &   $2.05 \times 10^{9}$     \\
  1.37   & 1.36972 &  0.00267802 &   0.00229552 &   9.719   &    9.854  &   $ 2.82 \times 10^{9}$ &   $4.91 \times 10^{9}$     \\
  1.38   & 1.37971 &  0.00241236 &   0.00177701 &   9.813    &   10.079  &   $ 4.13 \times 10^{9}$ &   $1.21 \times 10^{10}$     \\
  1.382  & 1.38170 &  0.00235540 &   0.00153125 &   9.834   &    10.210 &    $ 4.50 \times 10^{9}$  &      $2.03 \times 10^{10}$     \\
  \hline
  \end{tabular}                 
  \end{table*}

\begin{figure}
        \includegraphics[width=1.\columnwidth]{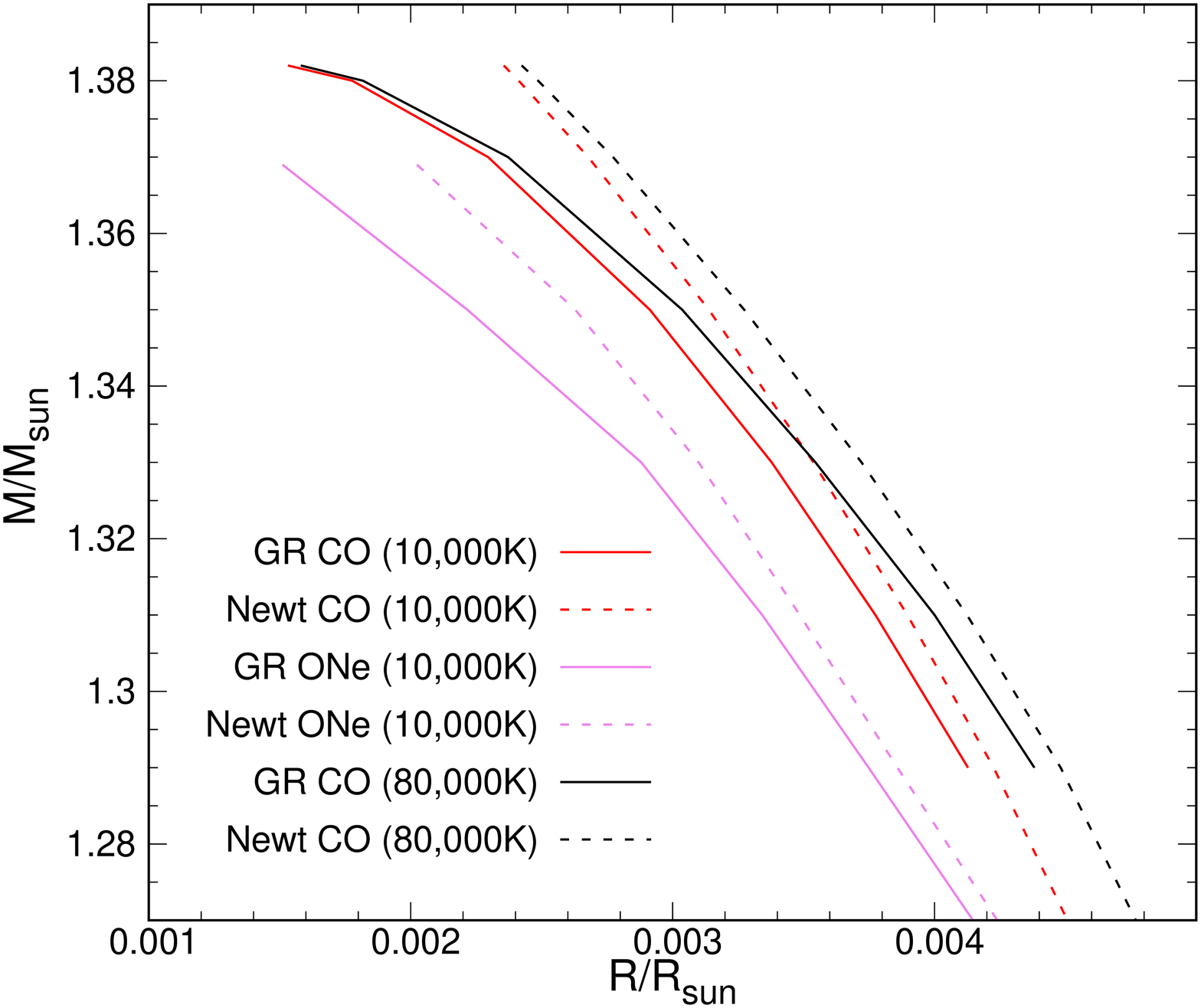}
        \caption{Gravitational mass versus stellar radius for our CO UMWD models considering and disregarding the effects of general relativity (solid and dashed lines, respectively). Models are shown at  $T_{ \text{eff}}=10\,000$ and $80\,000\,$K (red and black lines, respectively).  In addition, the mass-radius relations for ONe white dwarfs at $T_{  \text{eff}}=10\,000\,$K taken
from  \protect \cite{2022A&A...668A..58A}  are shown
with violet lines.} 
        \label{mr}
\end{figure}

\begin{figure*}
        \includegraphics[width=1.5\columnwidth]{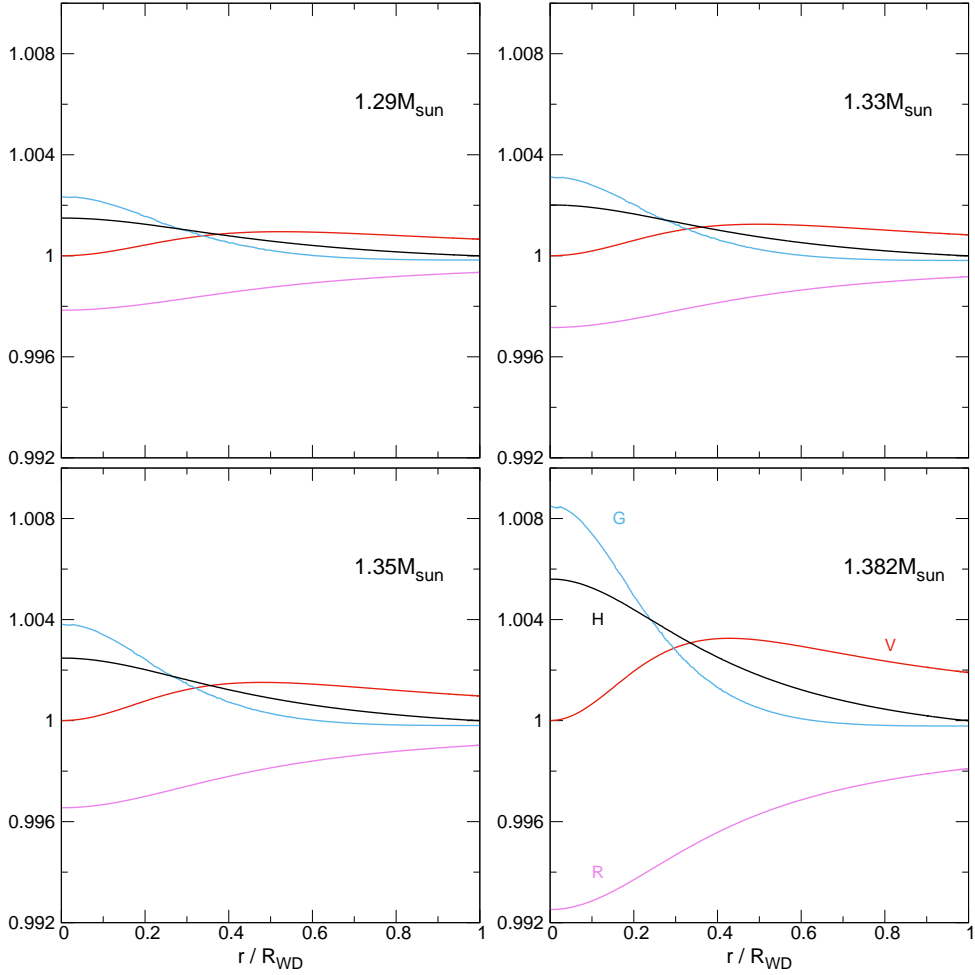}
        \caption{Run of general relativistic correction factors   $\mathscr{H}$, $\mathscr{G}$, $\mathscr{V}$,  and $\mathscr{R}$ (black, blue, red, and violet lines, respectively)
        for 1.29, 1.33, 1.35, and 1.382\ M$_{\sun}$ white dwarf
        models at log $L/L_{\sun}=-3$ in terms of the fractional radius. These factors turn to unity in the Newtonian  limit.} 
        \label{factors}
\end{figure*}

The rest of adopted input physics for our relativistic CO  white dwarf models
is the same as that  in \cite{2022MNRAS.511.5198C}, except that in the
present work we  did not explore the impact  of $^{22}$Ne sedimentation
on the  cooling of  white dwarfs.   We omitted  the energy  generation by
nuclear reactions  since these  are not happening  in our  white dwarf
models. We also disregarded neutrino cooling via the Urca process, despite the
fact that they are not entirely negligible in CO UMWDs,
see \cite{2021ApJ...916..119S}.  Hence,  the    cooling  times of our
relativistic sequences  may be somewhat  overestimated at
high and intermediate  luminosities. 
The energetics  resulting from  crystallization processes in  the core
was included as in \cite{2022MNRAS.511.5198C}, and 
takes into account the release
of latent heat  and the heat resulting from phase separation of carbon  and oxygen during crystallization.

We computed the full evolution  of constant rest-mass CO UMWDs  with masses of 1.29,  1.31, 1.33, 1.35, 1,37,  1.38, and 1.382 M$_{\sun}$.
Here,  the white dwarf mass means the  total rest mass or
baryonic mass  of the white  dwarf, which remains  constant throughout
the cooling process.  This mass should not be  confused with the total
gravitational mass,  $M_{\text{G}}$ (the value of  $m^t$ at the surface of the star),
that is, the stellar  mass that  would be
measured by  a distant  observer. $M_{\text{G}}$ changes  during the  course of
white dwarf evolution and turns out to be slightly lower than the total baryonic
mass of the white dwarf, see Table \ref{table1}.
As shown in \cite{2021A&A...646A..30A},  CO UMWDs
can  result  from  single  stellar  evolution  if  C  burning  on  the
asymptotic giant branch is avoided. In particular, for all the white
dwarf models we  employed the chemical profile of the 1.159  M$_{\sun}$
CO-core  hydrogen-rich   white  dwarf  resulting  from   the  complete
evolution of an initially 7.8\ M$_{\sun}$  model that avoids C-ignition, and
whose CO core slowly grows during  the super asymptotic giant
branch     as      a     result     of     reduced      wind     rates
\cite[see][]{2021A&A...646A..30A}.  This progenitor evolution  was computed
with    the    Monash-Mount    Stromlo   code    as    presented    in
\cite{gilpons2013,gilpons2018}.  The  set of  relativistic cooling sequences  of  CO
UMWDs presented  in this  work extends  the mass
range   of  the  CO   white    dwarf   sequences   already   computed   in
\cite{2022MNRAS.511.5198C}  within  the  framework  of  the  Newtonian
theory of stellar interiors.

\begin{figure}
        \centering
        \includegraphics[width=1.\columnwidth]{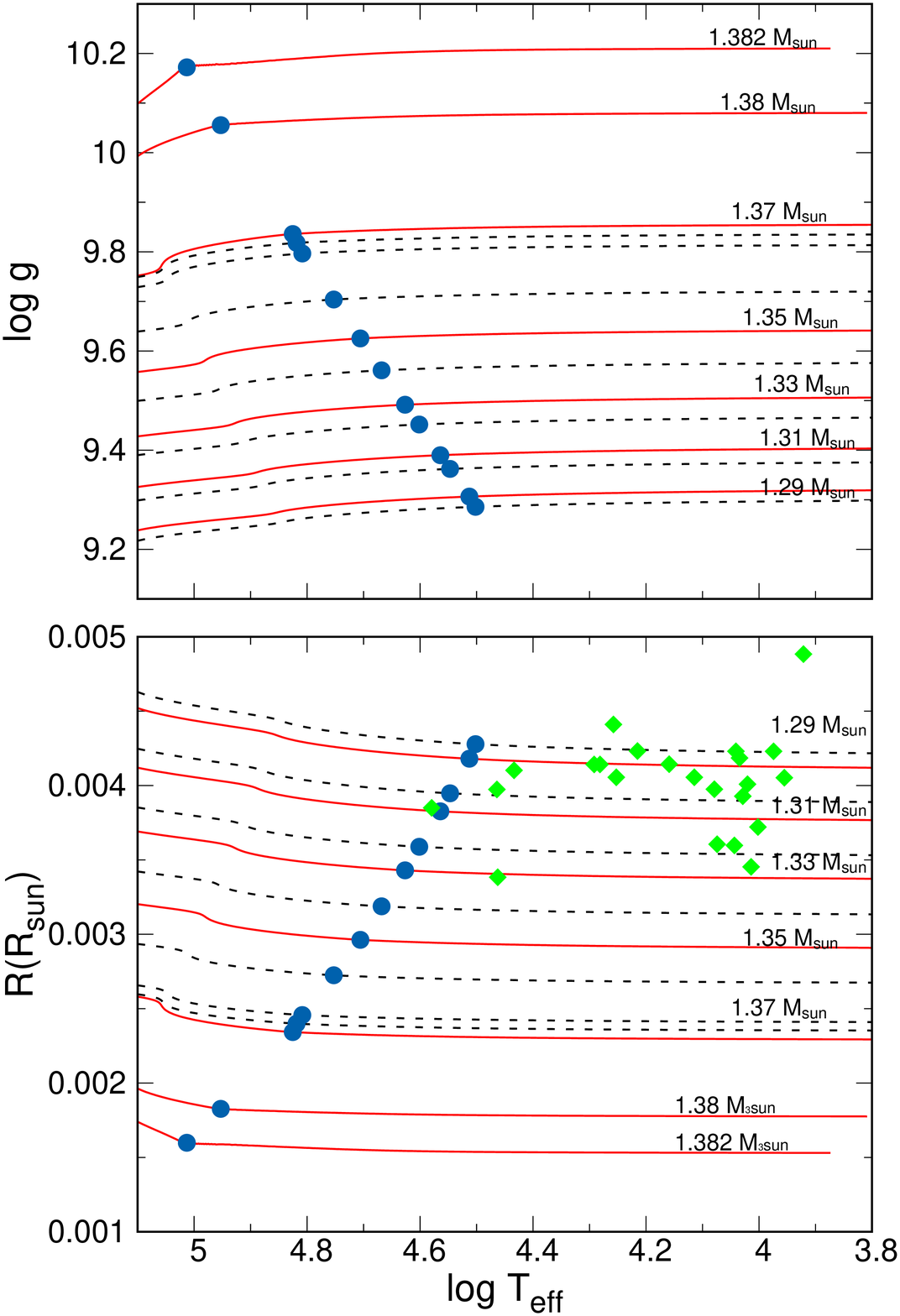}
        \caption{Surface gravity and stellar radius (in solar units) in terms
        of effective temperature for all of our
        sequences are displayed in the upper and bottom panels, respectively. Red solid and black dashed lines correspond to the general relativity  and Newtonian cases, respectively. From bottom (top) to top (bottom), curves in the upper (bottom) panel  correspond to
        1.29, 1.31, 1.33, 1.35, 1.37, 1.38,  and 1.382 M$_{\sun}$   CO white dwarf cooling sequences. Blue filled circles denote the onset of core crystallization in each sequence.
          The most massive white dwarfs in the solar neighbourhood analysed in \protect \cite{2021MNRAS.503.5397K} are displayed using green filled diamonds.
          }  
        \label{grav-surface}
\end{figure}

\begin{figure}
        \centering
        \includegraphics[width=1.\columnwidth]{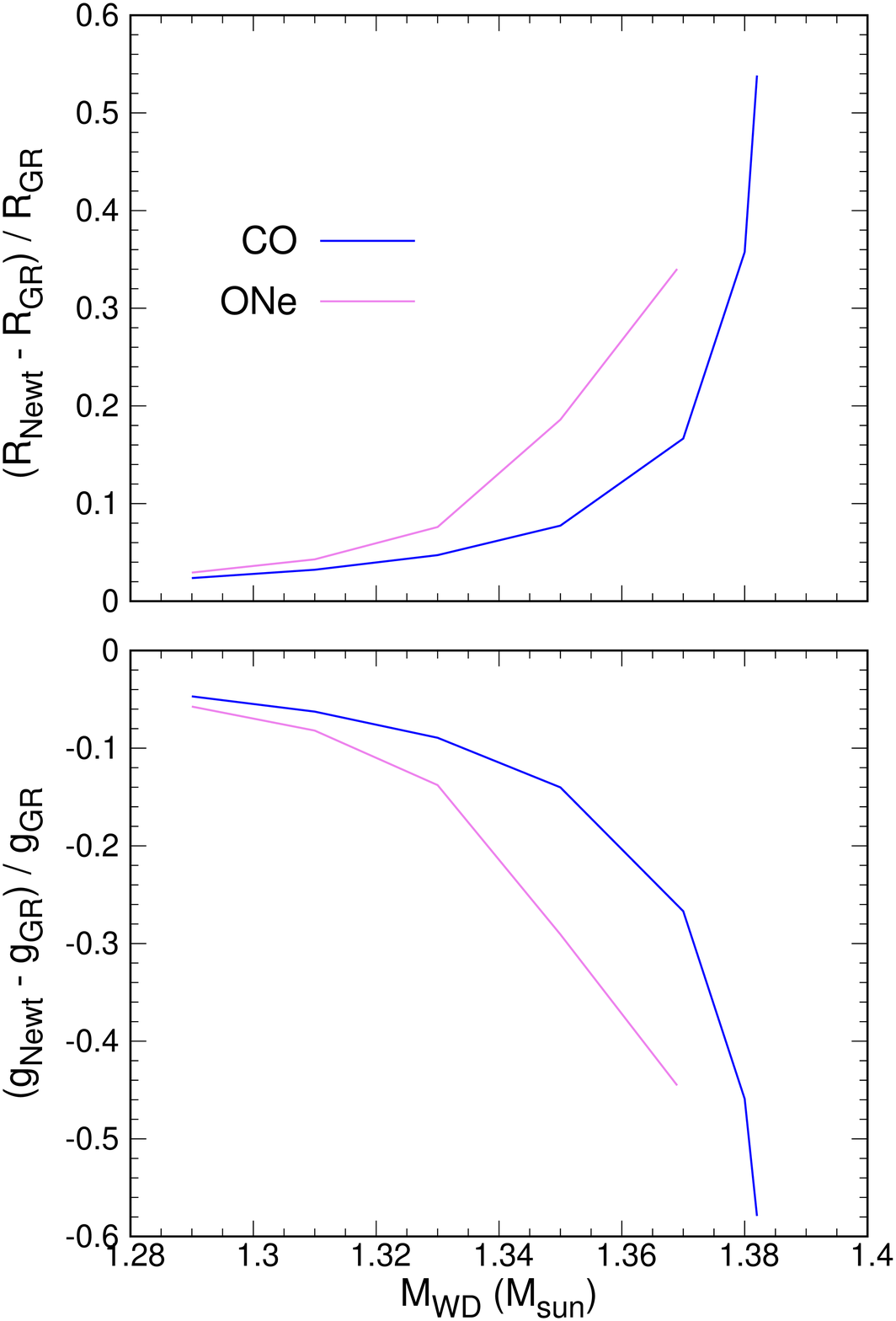}
        \caption{Upper panel: Relative difference between Newtonian stellar radius
          and general relativistic stellar radius in terms of the stellar mass.
          Results are for our CO and ONe UMWD models (blue
          and violet lines, respectively) at  $T_{  \text{eff}}=10\,000\,$K. Bottom panel: same
        as upper panel but for the surface gravity.} 
        \label{delta_gr}
\end{figure}

\begin{figure}
        \centering
        \includegraphics[width=1.\columnwidth]{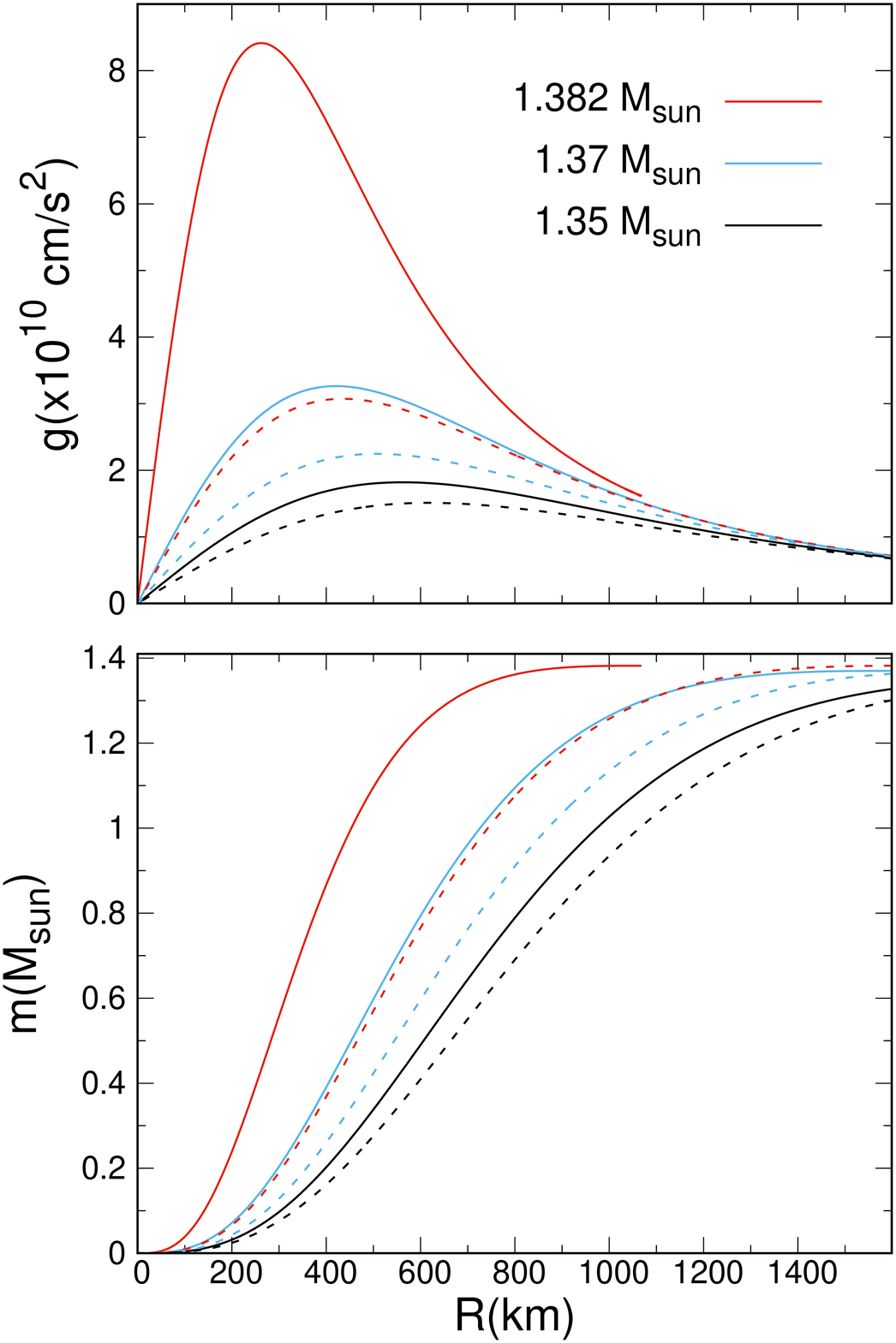}
        \caption{Gravitational field (upper) and rest mass $m$ (bottom panel)  for
          the general relativity and Newtonian cases (solid and dashed lines, respectively)
          in terms of radial coordinate 
          for 1.382, 1.37, and 1.35 M$_{\sun}$  CO white dwarf models at
          $T_{  \text{eff}}=11\,000\,$K (red, blue, and black lines, respectively).} 
        \label{gravity}
\end{figure}

\begin{figure}
        \centering
        \includegraphics[width=1.\columnwidth]{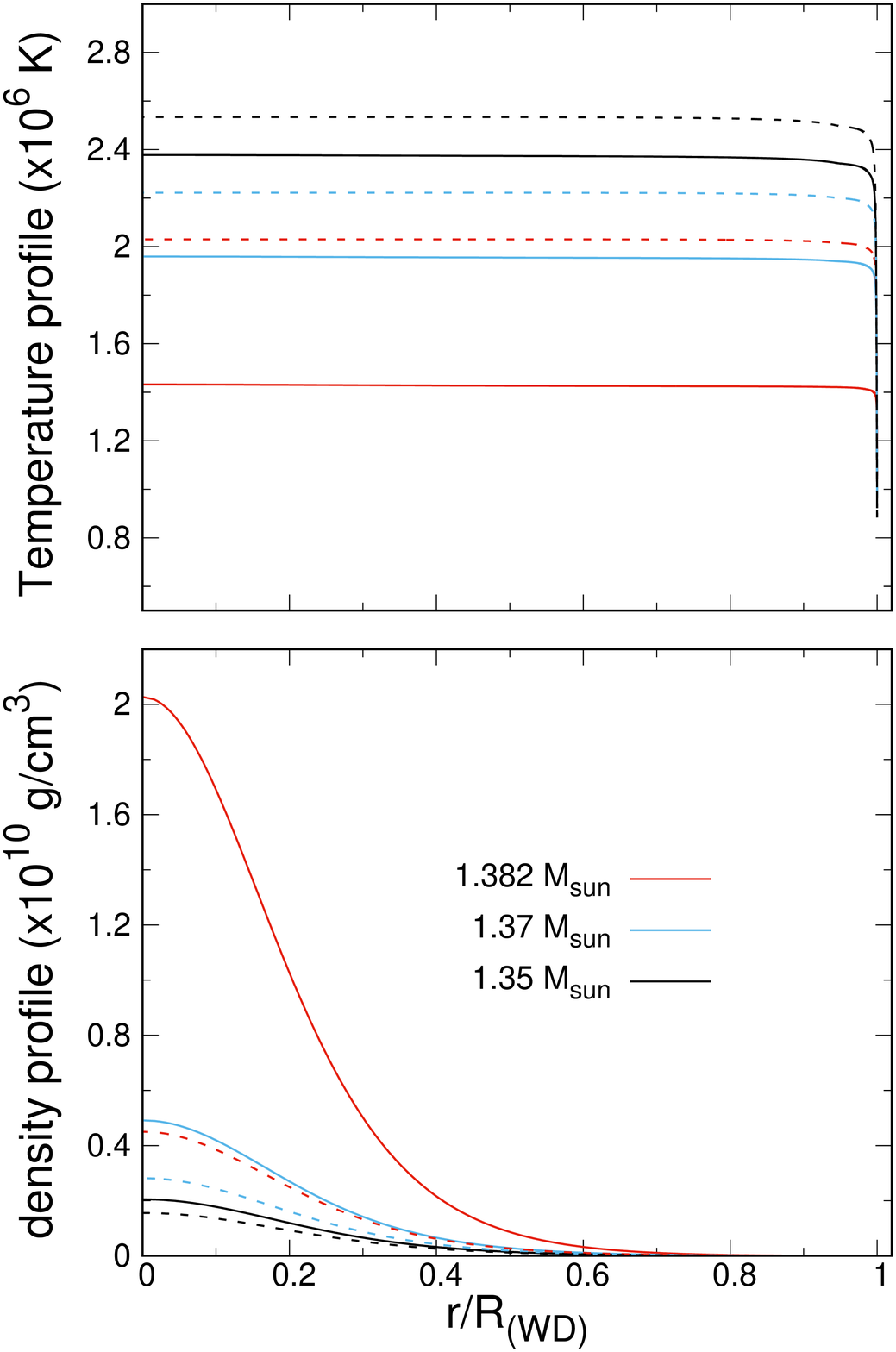}
        \caption{Temperature (upper panel) and density of rest mass (bottom panel)
          for the general relativity and Newtonian cases (solid and dashed lines,
          respectively)  in terms of radial coordinate for the same CO white
          dwarf models as depicted in Fig.\ref{gravity}.} 
        \label{trho}
\end{figure}

\section{Impact of general relativity on the evolution of massive CO white dwarfs}
\label{results}

We describe  now the  main consequences of  general relativity  on the
properties of CO UMWDs. In Fig.\,\ref{mr} 
 we show
the  resulting  mass-radius  relation at  two different  effective
temperatures.   Specifically, the  gravitational mass  is depicted  in
terms  of the  stellar radius  for  our CO  UMWD
models considering and disregarding  the effects of general relativity
(solid  and dashed  lines, respectively)  at $T_{\text{eff}}=10\,000$ and
$80\,000\,$K,  red and  black lines,  respectively. We mention that the
gravitational mass is lower by 0.02$\%$ than the baryonic stellar mass, see Table \ref{table1}. At  a given  mass, the stellar  radius is  smaller in  the case  that  general relativity
effects are taken into account, 
particularly
for highest stellar masses where  general relativity plays a  major role. 
In both cases,  
the impact  of finite  temperature  on the  stellar radius  is
noticeable only at lower stellar  masses and appears to show a somewhat higher 
variation with the stellar mass when the general relativity effects are introduced.  
For comparison purposes, we also
show In Fig.\,\ref{mr} the  mass-radius  relations  for ONe  white  dwarfs  at  $T_{ \text{eff}}=10\,000\,$K
taken from  \cite{2022A&A...668A..58A}  (violet  lines). Because of the
stronger Coulomb  interaction in ONe  cores, the stellar  radius results
smaller in ONe-core white dwarfs than in CO-core ones with the same mass.

In  our calculations,  CO white
dwarfs  more  massive  than 1.382  M$_{\sun}$  become  gravitationally
unstable with respect to  general relativity  effects, being this limit higher than the 1.369\  M$_{\sun}$
we found for ONe white  dwarfs, \cite{2022A&A...668A..58A}. This  happens at
a given finite central density somewhat larger than  $2 \times 10^{10}$ g cm$^{-3}$ (see
Table \ref{table1}). This limiting mass is  in  agreement with the findings of
zero-temperature
predictions by  \cite{2011PhRvD..84h4007R}   for 
pure-oxygen and pure-carbon   white      dwarfs
(1.380 and 1.386 M$_{\sun}$, respectively)  and with  \cite{2017RAA....17...61M}
for white dwarfs composed of oxygen (1.385 M$_{\sun}$) and carbon (1.391 M$_{\sun}$).
We note that the  central  density  of our 1.382\ M$_{\sun}$  white dwarf  model
in  the general relativity case  is near the  density threshold for
inverse $\beta-$decays for oxygen composition, $1.90 \times 10^{10}$ g cm$^{-3}$,
see \cite{2017RAA....17...61M}. For carbon composition, the corresponding threshold
density is  $3.9 \times 10^{10}$ g cm$^{-3}$, so  the central density is limited by
general relativity rather than electron captures in carbon white dwarfs. Consequently, the fact that we have not considered eventual instabilities  against the  inverse
$\beta-$decays should not affect the reliability of our models, at least up to white dwarf masses of 1.380 M$_{\sun}$.

The major  role played  by general  relativity for  increasing stellar
masses is  expected from  the dependence  of the  general relativistic
correction  factors with  the  stellar mass,  as  illustrated by  Fig.
\ref{factors}.  This   figure  displays   $\mathscr{H}$, $\mathscr{G}$, 
$\mathscr{V}$,  and $\mathscr{R}$ (black,  blue, red,  and violet lines,  respectively) in
terms  of  the  fractional  radius  for  the  1.29,  1.33,  1.35,  and
1.382\ M$_{\sun}$  white   dwarf  models  at  log   $L/L_{\sun}=-3$.  As
mentioned, these factors  are equal to one in the Newtonian limit,  and have a
slight  dependence on  the effective  temperature. Their  behaviour is
related to curvature effects, as well as to the fact that the pressure
and the  internal energy  appear as  a source  of gravity  in general
relativity,  which implies  that both  density and  pressure gradients
need to be  steeper than in Newtonian gravity  to maintain hydrostatic
equilibrium. This causes the  factors $\mathscr{G}$ and $\mathscr{H}$,
which depend on  density and pressure, to increase  towards the center
of the  star. The relativistic  factor $\mathscr{V}$, a  correction to
the volume,  is equal to one at  the center of the  star where the  volume is
zero  and  attains  a  maximum  value  at  some  inner  point  in  the
star. Towards the surface of  the star, $\mathscr{G}$ and  $\mathscr{H}$
tend to $M_{\text {G}}/M_{\text{WD}}$ and 1, respectively. We mention that the behaviour of
the general relativistic correction  factors depicted in   Fig. \ref{factors}
qualitatively resemble those exhibited by the ONe UMWDs
\citep[see][]{2022A&A...668A..58A}.

In  Fig. \ref{grav-surface} we illustrate the surface  gravity and the stellar radius
in terms of the effective temperature
for  all of  our sequences  for the  general relativity  and Newtonian
cases, using solid and dashed lines, respectively. The gravitational field  for the general relativistic sequences 
corresponds to the gravitational field  as measured far from the star, and is given by

\begin{equation}
g^{\text{GR}}= \frac{G m}{r^2} \mathscr{G} \mathscr{V}^2 \ .
\label{grav_sup}
\end{equation}

Clearly,   the surface  gravity  and stellar  radius are  
markedly affected by   general relativity, particularly for the most massive
sequences. We note  that not considering the general relativity
effects would yield a stellar mass value about 0.015\ M$_{\sun}$  larger
for cool white dwarfs with  measured surface gravities of log g $\approx$ 9.8.
The  photometric   measurements  of
\cite{2021MNRAS.503.5397K} for  the radius of the UMWDs
 in the solar neighbourhood are  also plotted in this figure. According to our CO ultra-massive relativistic sequences, the most massive white dwarfs of this sample have masses near 1.33\ M$_{\sun}$. Such value turns out  to be
 1.34\ M$_{\sun}$ if Newtonian sequences were used instead.

In Fig.  \ref{delta_gr} we show  the relative difference in stellar radius and gravity, upper and lower panel, respectively, between Newtonian and general relativistic predictions
versus  the stellar mass for our CO UMWD models at  $T_{ \text{eff}}=10\,000\,$K.
For comparison purposes we also show the predictions for the relativistic ONe white dwarfs
from \cite{2022A&A...668A..58A} (violet lines). Clearly,
differences increase markedly as the stellar mass approaches our maximum mass value,
1.382\ M$_{\sun}$, for which the relative difference in radius (taking as a reference the relativistic model) is about $55\%$.
The stellar radius at this stellar mass results 1070 km in the general relativity case
(see  Table  \ref{table1}), which is nearly $40\%$ smaller than the moon radius. For stellar masses lower than $\approx$ 1.29\ M$_{\sun}$,
differences in the stellar radius resulting from general relativistic corrections  become lower than $2\%$.

The onset of  core crystallization   in each  sequence depicted in Fig. \ref{grav-surface}
is marked with blue  filled circles. Because of the denser and cooler cores
of the general relativistic white dwarfs (see later in this section),
core crystallization at a given stellar mass begins at  higher effective temperatures
than in their Newtonian counterparts. This is particularly evident
for the 1.382\ M$_{\sun}$ white dwarf sequence. We note that for this stellar mass,
general relativity predicts that a $70\%$ of the white dwarf mass has
crystallised  by the time the onset of core crystallization takes place in the
Newtonian case. The marked reduction in the radius at high effective temperatures before
the onset of crystallization is due to the dominance of neutrino cooling at those
evolutionary stages.

The run of the inner  gravitational field and rest mass  versus radial  coordinate
for the  1.382, 1.37, and 1.35\ M$_{\sun}$  CO white dwarf models at
$T_{ \text{eff}}=10\,000\,$K is shown in Fig. \ref{gravity}  (upper and bottom panel,
respectively) for
the general relativity and Newtonian cases (solid and dashed lines, respectively).
In the case of the most massive model,  general   relativity  effects   strongly
impact  the  stellar structure. We note in particular the noticeable differences in the gravitational field, as comparted with Newtonian predictions.
The impact is
also evident toward  lower stellar masses, albeit to a  lesser extent. In  Fig. \ref{trho} we show the run of temperature  and density of rest mass
for the general relativity and Newtonian cases
(solid and dashed lines, respectively)  in terms of radial coordinate for the same
CO white dwarf models as depicted in Fig.\ref{gravity}. We note the 
larger inner densities  in the general relativity models, as compared to the Newtonian case, particularly for the
most massive white dwarf model.  In addition, appreciable differences arises in the internal
temperature profile. Models with  1.382\ M$_{\sun}$  (1.35\ M$_{\sun}$)  computed considering general relativity  have central temperatures about $30\%$  ($6\%$) cooler than their Newtonian counterparts during most of cooling stages.

An important observational constraint to the mass-radius relation of white dwarfs is
provided by the measurement of the gravitational redshift, related to
the fact that photons lose energy
when escaping from the surface of a white dwarf. The resulting change in the photon wavelength
($\lambda$) is related to the recession velocity by, see \cite{2020ApJ...899..146C}:

\begin{equation}
v_{\rm g}= c\ \frac{\Delta \lambda}{\lambda},
\label{velocity}
\end{equation}

\noindent where

\begin{equation}
\frac{\Delta \lambda}{\lambda}= \mathscr{V} -1.
\label{deltalambda}
\end{equation}

\begin{figure}
        \centering
        \includegraphics[width=1.\columnwidth]{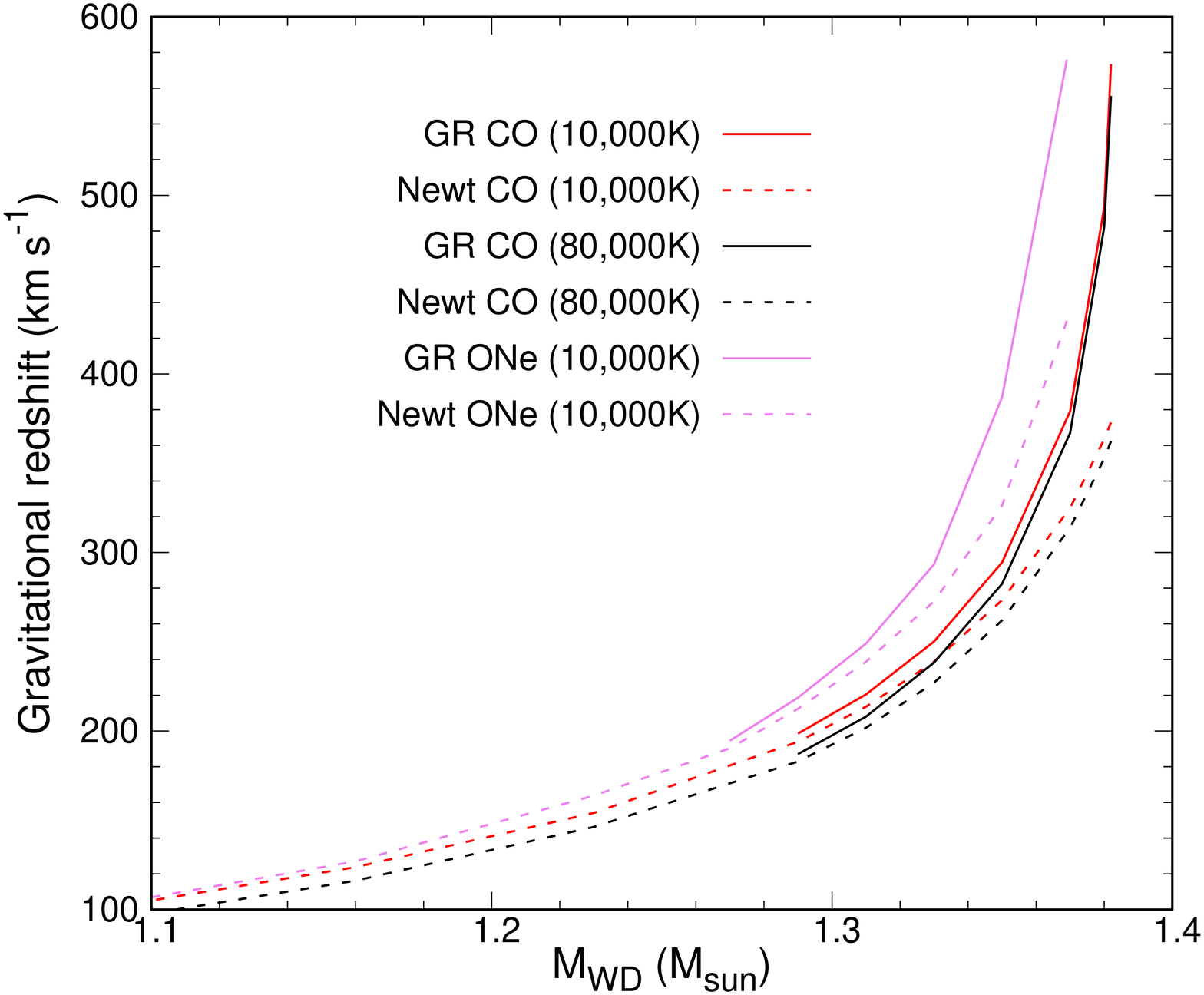}
        \caption{Gravitational redshift versus stellar mass for our CO UMWD models considering and disregarding the effects of general relativity (solid and dashed lines, respectively). Models are shown at  $T_{  \text{eff}}=10\,000$ and $80\,000\,$K (red and black lines, respectively).  In addition, the gravitational redshift for ONe white dwarfs at $T_{  \text{eff}}=10\,000\,$K is shown
with violet lines.} 
        \label{redshift}
\end{figure}

\begin{figure}
        \centering
        \includegraphics[width=1.\columnwidth]{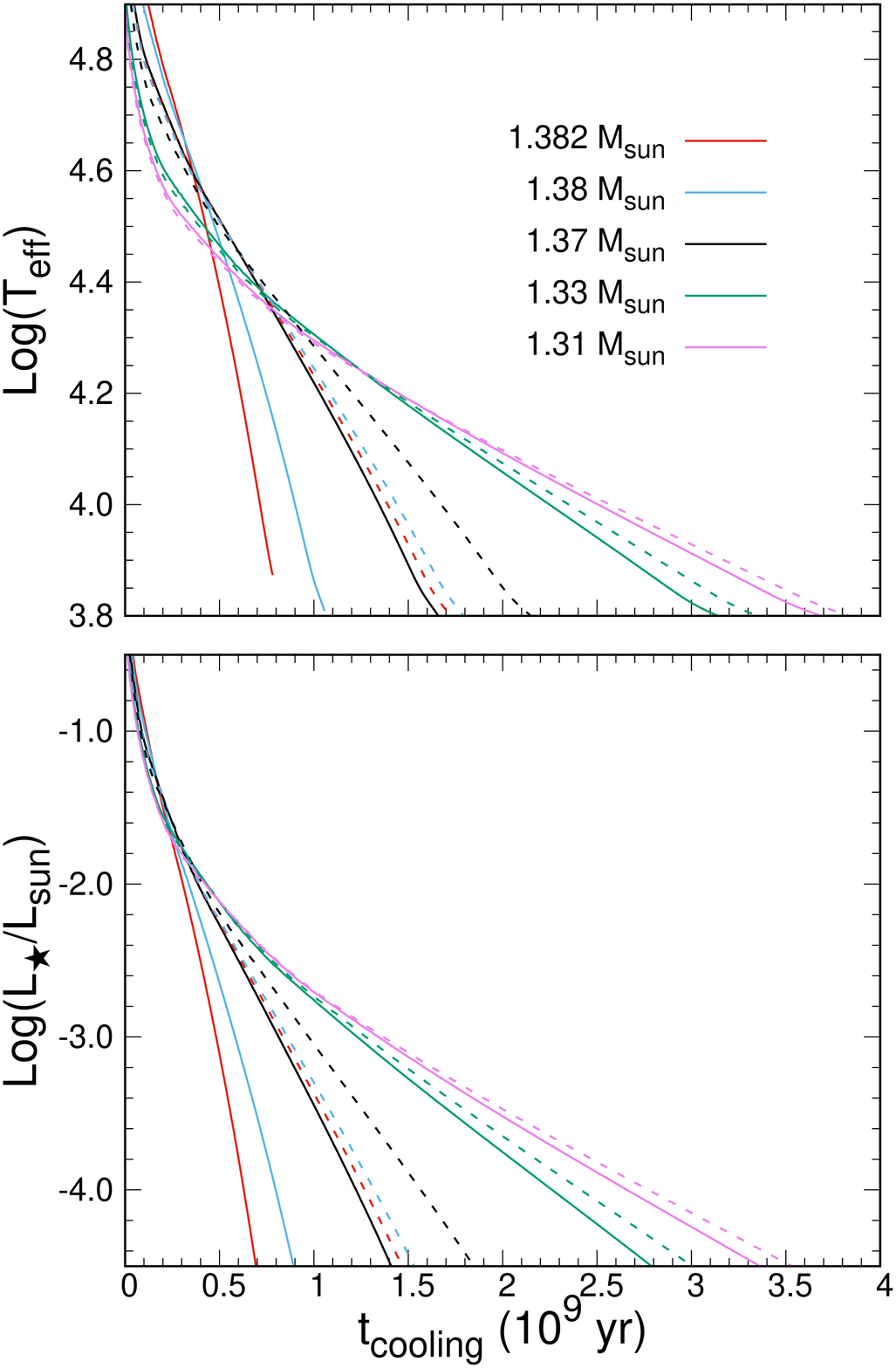}
        \caption{Effective temperature and surface luminosity (upper and bottom panels) versus the cooling times for our  1.31,  1.33, 1.37, 1,38, and 1.382\ M$_{\sun}$
          CO white dwarf sequences.  
        Solid (dashed) lines correspond to the general relativity (Newtonian) cases.
        Cooling time is counted from the time of white dwarf formation.} 
        \label{age}
\end{figure}

\begin{figure*}
    \includegraphics[width=1.8\columnwidth]{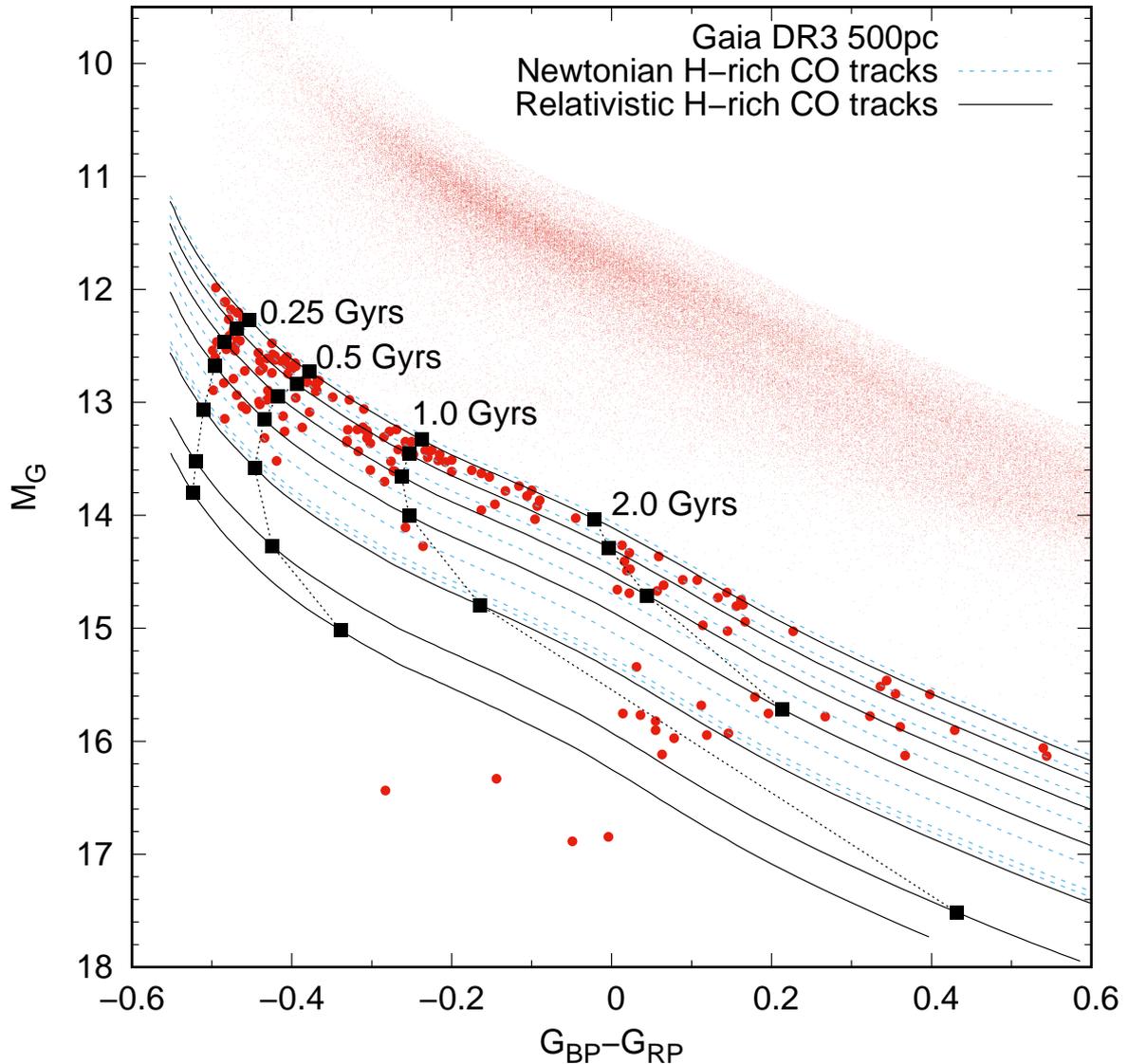}
        \caption{{\it Gaia} colour-magnitude diagram.  Newtonian and general relativistic CO-core cooling sequences are displayed using dashed blue and solid black lines, respectively. Their rest masses are, from top to bottom, 1.29, 1.31, 1.33, 1.35, 1.37, 1.38 and 1.382\ M$_{\sun}$. The {\it Gaia} DR3 white dwarf population within 500 pc from Torres et al. (2023; submitted), is displayed using red dots. Among these objects, we highlighted the UMWDs 
        located below the  1.29 M$_{\sun}$ cooling track
        using red filled circles. Dotted black lines display isochrones of   0.25, 0.5, 1, and 2 Gyrs.
       } 
        \label{gaia}
\end{figure*}

Here,  $\mathscr{V}$  is given  by  assessing  Eq. \ref{eq_V}  at  the
stellar surface.  Since the gravitational redshift  depends directly on
the mass and  radius of the white dwarf, we  expect this quantity
to be altered  by general relativity in massive  white dwarfs.   This is
borne  out by  Fig.  \ref{redshift}, which  depicts the  gravitational
redshift in  terms of the stellar  mass for our CO  UMWD  models  considering and  disregarding  the  effects of  general
relativity  (solid   and  dashed   lines,  respectively)   at  $T_{\text{eff}}=10\,000$ and  $80\,000\,$K (red and black  lines, respectively). For
comparison purposes, predictions  for  ONe white dwarfs
    from \cite{2022A&A...668A..58A} 
    at  $T_{  \text{eff}}=10\,000\,$K
    are also shown. For  the  most massive  white dwarf  models, the  inclusion of
general  relativity effects  leads to  a gravitational  redshift a
factor 1.5 larger than predicted by the Newtonian white dwarf  models (differences in the gravitational
redshift are about 200 km s$^{-1}$). This result was
expected since $v_{\text{g}}$ varies  essentially as $\approx M/R$, and the
stellar  radius  is  markedly   smaller  when  general  relativity  is
taken into account.   For masses below 1.29\ M$_{\sun}$,  differences  in the  gravitational
redshift are below 2-3 km s$^{-1}$. Because ONe white dwarfs are more compact than
CO ones, a larger gravitational redshift is expected in these stars.

\begin{figure}
        \centering
        \includegraphics[clip,width=240pt]{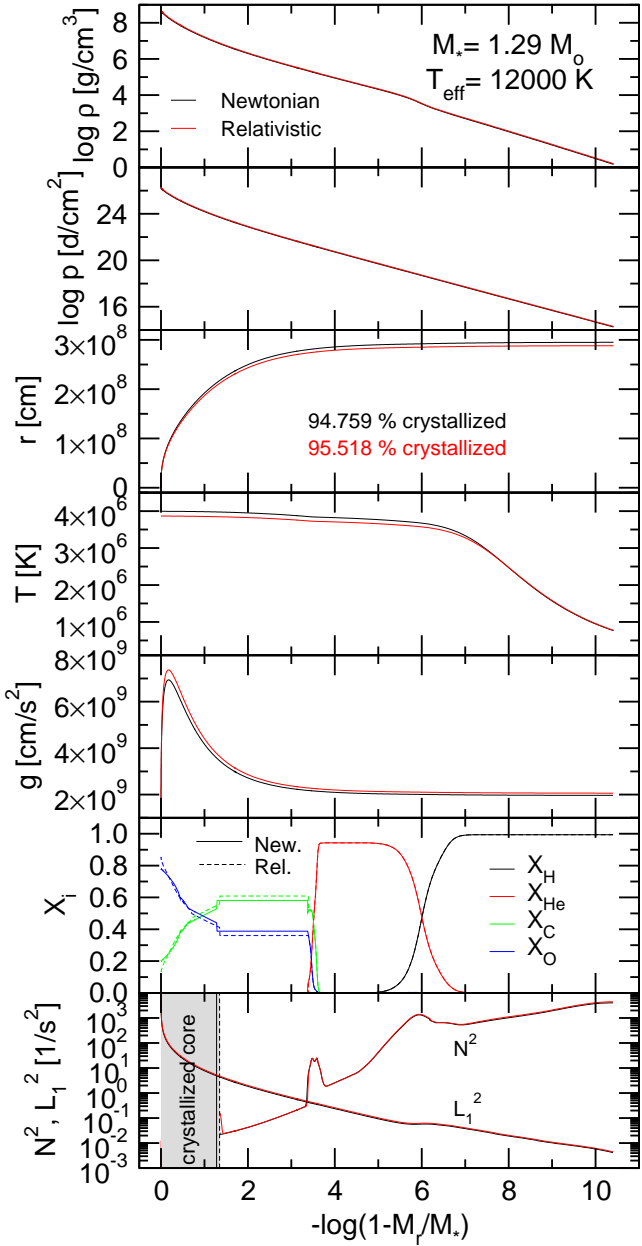}
        \caption{Newtonian (black curves) and relativistic (red curves) properties of a template UMWD model characterised by  $M_{\star}= 1.29\ \text{M}_{\sun}$ and $T_{  \text{eff}}\sim 12\,000$ K. From top to bottom, we depict the run of the logarithm of the density ($\rho$), the logarithm of pressure ($p$), the radius $r$, the temperature ($T$), the gravity ($g$), the internal chemical profiles ($X_i$), and the squared critical Brunt-V\"ais\"al\"a and Lamb ($\ell= 1$) frequencies, in terms of the outer mass fraction coordinate. The grey region in the bottom panel represents the crystallised core, in which the $g$ mode-pulsations are inhibited.  In the plot 
        of the chemical profiles and location of the crystallization front, solid (dashed) curves correspond to the Newtonian (relativistic) case.}
        \label{pulsa-1.29}
\end{figure}

\begin{figure}
        \centering
        \includegraphics[clip,width=240pt]{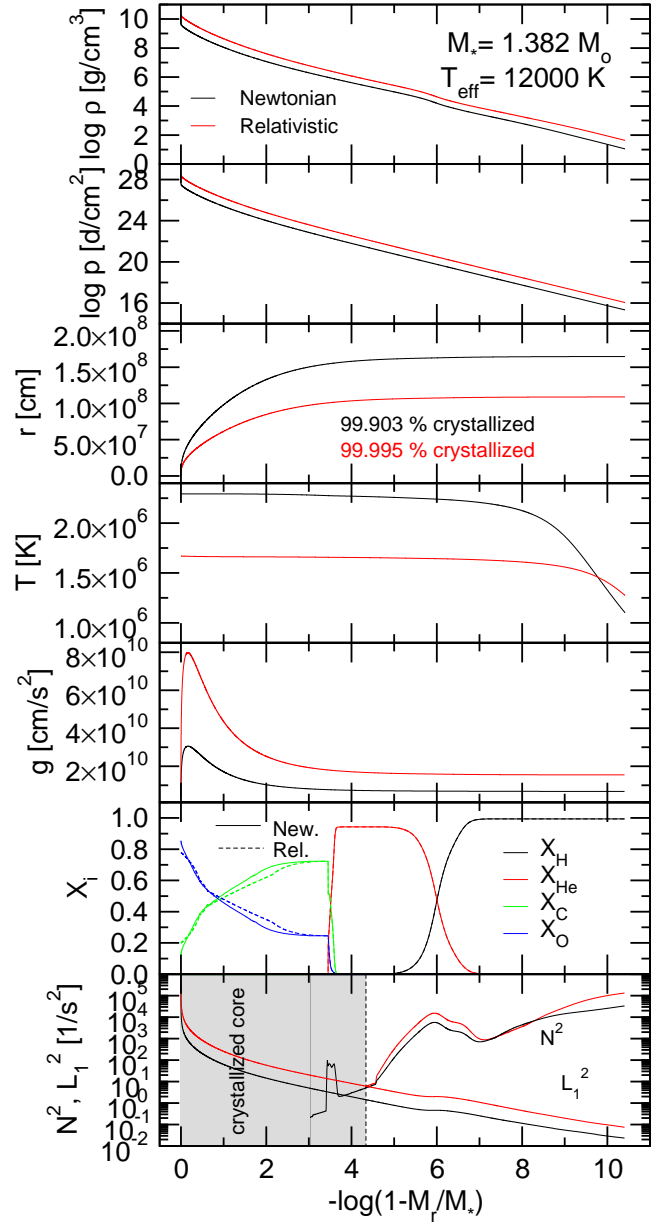}
        \caption{Same as in Fig. \ref{pulsa-1.29}, but for the case of a UMWD model with $M_{\star}= 1.382\ {\text{M}}_{\sun}$
        .} 
        \label{pulsa-1.382}
\end{figure}

In Fig. \ref{age} we illustrate the cooling times  of our CO  UMWDs  for  the  general
relativity   and    Newtonian   cases,   with solid   and    dashed   lines,
respectively. The  cooling times are set  to zero at the  beginning of the
cooling  tracks  at  very high  effective  temperatures. 
As in the case of ONe white dwarfs we studied in \cite{2022A&A...668A..58A}, the cooling
behaviour is substantially  altered by  general relativity effects. In the case of the
most massive white dwarf, general   relativity   causes
CO UMWDs to evolve  faster than in the Newtonian
case  at advanced  stages  of evolution. In particular,  the
1.382\ M$_{\sun}$  relativistic sequence requires half the time to reach
the low luminosity stages than needed in the Newtonian case. The  trend in
the cooling behaviour  is reversed at
earlier  stages  of evolution,  where  white  dwarfs computed  in  the
general   relativity  case   evolve   more slowly   than  their   Newtonian
counterparts, see \cite{2022A&A...668A..58A}  for a similar behaviour and explanation.

\subsection{Relativistic UMWD {\it Gaia} candidates}

The {\it Gaia} mission has provided an unprecedented increase in both the quality and the number of known white dwarfs in our solar neighbourhood.  
Recently, Torres et al. (2023; submitted) analysed a sample of nearly
90\,000 white dwarfs within 500 pc from the Sun obtained from {\it Gaia}
DR3 data. The sample has been selected  applying the same criteria as in
\citep{Jimenez2023} but extended up to 500 pc. Although completeness is not
guaranteed,  the applied selection criteria ensure good quality data
(i.e., photometric and astrometric errors below 10$\%$).
In Fig. \ref{gaia} we show the {\it Gaia} Hertzsprung-Russell diagram for the entire sample (red dots) of the white dwarf population. Superimposed are the  {\it Gaia} magnitudes\footnote{Sloan  Digital Sky  Survey, Pan-STARRS  and other  passbands are  also
available upon  request using the  non-grey model
atmospheres                                                         of
\cite{2010MmSAI..81..921K} and \cite{2019A&A...628A.102K}.} tracks  for  our  relativistic and  Newtonian CO cooling sequences (solid black and dashed blue lines, respectively).  Isochrones of   
0.25, 0.5, 1,  and 2 Gyr for  our relativistic sequences are  also shown  with dotted black  lines.   An  inspection of  the {\it  Gaia}  colour-magnitude  diagram reveals an important set of objects with masses higher than 1.29\ M$_{\sun}$ (red filled circles) that can be considered as reliable UMWD candidates. Among these candidates we clearly identify those belonging to the so-called faint-blue branch ($M_G>15$) that deserve a separate analysis \citep[see for instance][]{2022RNAAS...6...36S,2022A&A...668A..58A,2022ApJ...934...36B}. For the rest of objects, around 25 of them lay in the colour-magnitude region corresponding to masses larger than 1.33\ M$_{\sun}$. It is beyond the purpose of the present work to fully analyse these objects, as it would require higher accurate observations; however, they can be presented as reasonable UMWD candidates of our solar neighbourhood for which general relativistic effects should be taken into account.
In the case of the most massive white dwarfs, general relativity effects lead to  white dwarf sequences that are markedly 
fainter than Newtonian sequences with the  same mass. Thus, the effects of 
general relativity  must be carefully taken into account when determining the mass
and stellar properties  of such  massive white  dwarfs through {\it
  Gaia}  photometry. Not  considering such  effects would  lead to  an
overestimation  of their  mass and  an incorrect  estimation of  their
cooling times.

\section{Pulsational properties of relativistic white dwarf models}
\label{pulsational}

Given the continuous arrival of large amounts of photometric data of pulsating white dwarfs from space missions such as the ended {\it Kepler}/{\it K2} program and the ongoing 
{\it TESS} mission, we believe to be worthwhile to assess the pulsational properties of our relativistic UMWDs. 
As a first step, we limit ourselves to calculate Newtonian pulsations of non-radial $g$ modes on fully relativistic CO UMWD  models, similar to what was done in the preliminary calculations of \cite{2023arXiv230204100C} for ONe UMWDs. The assessment of the combined effects of considering relativistic ONe- and CO-core UMWD models {\it along with} fully relativistic $g$-mode pulsations, will be the focus of a future publication (C\'orsico et al. 2023, in preparation).

We have selected two pairs of template CO white dwarf models with $M_{\star}= 1.29\ \text{M}_{\sun}$ and $M_{\star}= 1.382\ {\text{M}}_{\sun}$, each couple of models considering and disregarding the effects of general relativity. These models have $T_{  \text{eff}}= 12\,000$ K, an effective temperature representative of the instability strip of the ZZ Ceti variables. We have computed non-radial $g$ modes with periods between 50 s and 1500 s for these template models using an adiabatic version of the {\tt LP-PUL} pulsation code \citep{2006A&A...454..863C} that assumes the "hard-sphere" boundary conditions for the eigenfunctions at the edge of the solid core of crystallizing white dwarf models \citep{1999ApJ...526..976M}.
We show in Fig. \ref{pulsa-1.29} the logarithm of the density,  pressure, the stellar radius, the temperature, the gravity, the internal chemical profiles, and the squared Brunt-V\"ais\"al\"a and Lamb ($\ell= 1$) frequencies, in terms of  logarithm of the outer mass fraction coordinate, corresponding 
to the template UMWD model characterised by  $M_{\star}= 1.29 \text{M}_{\sun}$ and $T_{ \text{eff}}\sim 12\,000$ K. Black curves correspond to the Newtonian case, while red curves are display the fully relativistic case. Clearly, general relativity induces smaller radii and temperatures,  and larger gravities, as compared with the Newtonian case. The effect on the density and pressure profiles is hardly visible, but non negligible.
Also, no appreciable differences arise in the crystallised mass from the two models. Finally, the effect of considering Newtonian or relativistic gravitation has very little impact on the chemical composition profiles and on the Brunt-V\"ais\"al\"a and Lamb frequencies.

The situation is completely different when we consider  more massive models, as can be seen in Fig. \ref{pulsa-1.382}, that show the same quantities as in Fig. \ref{pulsa-1.29}, but for the case of the template models with $M_{\star}= 1.382\ {\text{M}}_{\sun}$. As we described previously, in this case, there is a very large variation in the stellar radius, gravity and temperature when relativistic effects are taken into account. The change in density and pressure is now noticeable (note that these quantities are plotted on a logarithmic scale). The strong impact of considering relativistic or Newtonian gravity is reflected in major differences in the Brunt V\"ais\"al\"a and Lamb frequency profiles (see the lower panels of Fig. \ref{pulsa-1.382}). In addition, we note that in the relativistic model, the CO/He interface is reached by the crystallization front, in contrast with the situation for the Newtonian model.

\begin{figure*}
        \centering
        \includegraphics[clip,width=450pt]{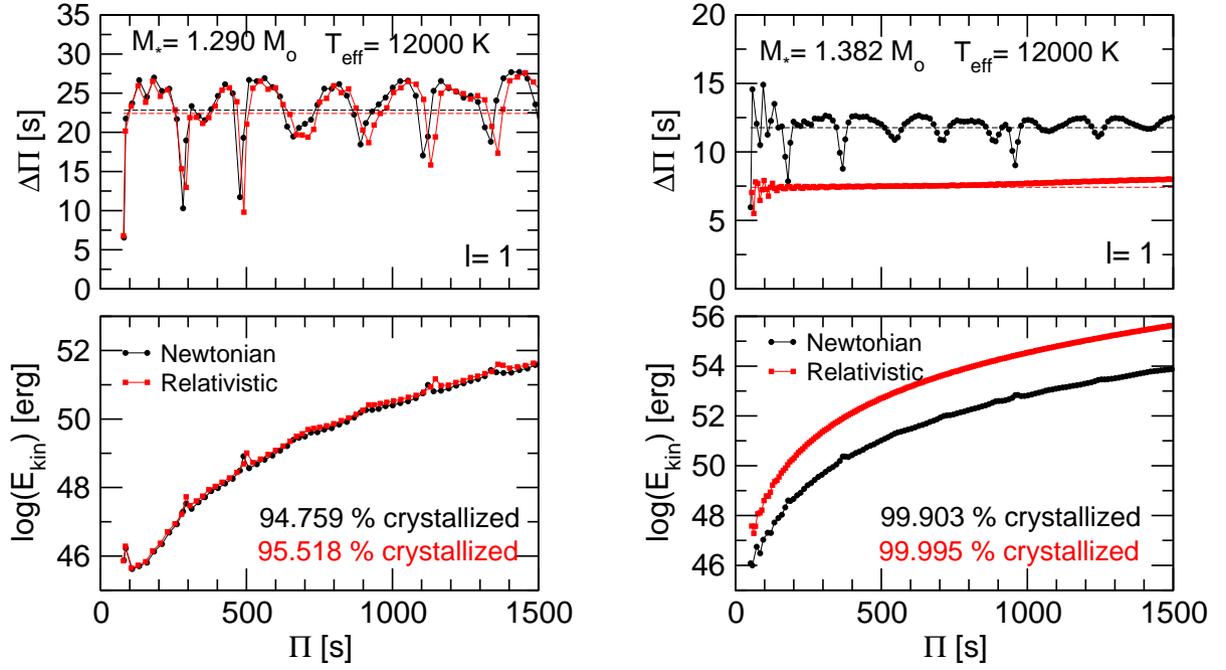}
        \caption{{\it Left panels:} the forward period spacing, $\Delta \Pi$ (upper panel), and the logarithm of the kinetic energy of oscillation, $E_{\text {kin}}$ (lower panel), in terms of the pulsation periods, $\Pi$, corresponding to the template UMWD model with $M_{\star}= 1.29\ \text{M}_{\sun}$ and $T_{  \text{eff}}\sim 12\,000$ K (see Fig. \ref{pulsa-1.29}). The results corresponding to the Newtonian case are displayed using black curves and symbols, while the relativistic results are shown with red curves and symbols. {\it Right panels:} same as in the left panels, but for the template UMWD model with $M_{\star}= 1.382\ {\text{M}}_{\sun}$ and $T_{  \text{eff}}\sim 12\,000$ K  (Fig. \ref{pulsa-1.382}).} 
        \label{prop-pulsa}
\end{figure*}

The impact on the pulsational properties of considering fully relativistic UMWD models in comparison with models assuming Newtonian gravity is shown in Fig. \ref{prop-pulsa} for the two values of stellar mass considered. The figure displays the forward period spacing (upper panels) and the kinetic energy of oscillation of modes (lower panels), in terms of the dipole $g$-mode  periods.  Not surprisingly, adopting general relativity does not have a significant impact neither on the period spacing nor on the kinetic energy of the modes for the case of the models with $M_{\star}= 1.29\ \text{M}_{\sun}$ (left panels). In contrast, there is a significant impact in the   $1.382\ {\text{M}}_{\sun}$ model (right panels). Indeed, the asymptotic period spacing (dashed horizontal lines) for the relativistic case is about five seconds shorter than for the Newtonian model. This is because the Brunt-V\"ais\"al\"a frequency is higher for the relativistic case (Fig. \ref{pulsa-1.382}), giving rise to a shorter period spacing. We can also notice a more featured distribution of $\Delta \Pi$ vs $\Pi$ for the Newtonian case, compared to the relativistic one. This is because in the Newtonian case, the propagation region (see the lower panel of Fig. \ref{pulsa-1.382}) includes two bumps which are associated with the CO/He and He/H chemical transitions, while in the relativistic case, the bump associated with the CO/He interface is embedded in the solid core, where the modes cannot propagate, and thus it does not impact the mode-trapping properties of the model. Finally, we note that the kinetic energy values are globally higher for the relativistic model. This is due to the fact that this model is much more compact than its Newtonian counterpart, which implies macroscopic movements of the fluid in a much denser environment, and therefore, with greater kinetic energies.

\section{Summary and conclusions}
\label{conclusions}

We presented the first set of evolutionary sequences for CO UMWD models   of  1.29, 1.31, 1.33, 1.35,  1.37, 1.38, and 1.382\ M$_{\sun}$, which  fully take into
  account the  effects of general  relativity on their  structural and evolutionary properties. These sequences extend the relativistic ONe UMWD sequences already computed in  \cite{2022A&A...668A..58A} to   models with a CO-core composition.
  CO UMWDs can result either from single or binary stellar evolution. In this work, initial models for our evolving sequences were extracted from the full single evolution 
  of an initially 7.8\ M$_{\sun}$  model that avoids C-ignition, and
whose CO core slowly grows during  the super asymptotic giant
branch     as      a     result     of     reduced      wind     rates
\cite[see][]{2021A&A...646A..30A}. This progenitor evolution  was computed
with    the    Monash-Mount    Stromlo   code    as    presented    in
\cite{gilpons2013,gilpons2018}. 
Calculations presented here were done with  La  Plata stellar evolution code, {\tt  LPCODE}, for which the standard stellar structure and evolution equations have been modified to include the
effects of general relativity.
For  comparison  purposes, additional sequences of identical initial models were  computed for the Newtonian gravity  case. Considering that the number of   UMWDs in the solar neighbourhood with  mass
determinations  beyond   about  $1.30\,  \text{M}_{\sun}$   has  substantially
increased  in recent  years \citep[e.g][]{2021MNRAS.503.5397K}, the calculations presented here are
timely to study the properties of such stars.  Also, from a sample of white dwarfs 
within 500\,pc from the Sun, we identified a number of reasonable UMWD candidates with masses larger than 1.33\ M$_{\sun}$  for which general relativistic effects should be taken into account.

We  provide  mass-radius relations for relativistic CO UMWD sequences, cooling times, and  magnitudes in {\it  Gaia} DR3,  
Sloan  Digital Sky  Survey, and Pan-STARRS  passbands  
using the  non-grey model atmospheres of
\cite{2010MmSAI..81..921K} and \cite{2019A&A...628A.102K}.
We find that CO white
dwarfs  more  massive  than 1.382\  M$_{\sun}$  become  gravitationally
unstable with respect to  general relativity  effects, being this limit higher than the limiting mass value of 1.369\  M$_{\sun}$ we found in 
\cite{2022A&A...668A..58A} for ONe  white dwarfs. As expected, the importance of general
relativistic effects increases as the white dwarf mass increases. As the stellar mass approaches the limiting mass value,
relative differences in radius (taking as a reference the relativistic model) reach up to
$55\%$. Hence, for
such massive white dwarfs, general relativity leads to white
dwarf sequences that are markedly fainter than Newtonian sequences with the same mass. Therefore, the effects of 
general relativity effects must be carefully taken into account when determining the mass
and stellar properties  of such  massive white  dwarfs.
We also find that the thermo-mechanical and evolutionary properties of the most massive  white dwarfs  are strongly affected by  general relativity effects. 
As in the case of ONe white dwarfs  studied in \cite{2022A&A...668A..58A}, we find that general
relativity strongly alters the evolutionary properties of the most massive CO UMWDs, leading, at advanced stages of evolution, to much shorter cooling times than in the Newtonian
case. We also find that for the  most massive  white dwarf  models, the  inclusion of
general  relativity effects  lead to  a gravitational  redshift a
factor 1.5 larger than predicted by the Newtonian white dwarf  models (differences in the gravitational
redshift are about 200 km s$^{-1}$).

The immediate prospect of detecting pulsating white dwarfs with masses above $\sim 1.30\  \text{M}_{\sun}$ \citep[see][]{2023arXiv230410330K}
motivated also us to assess in this work, for the first time, the impact of our new relativistic UMWD models
on the pulsational properties of ultra-massive ZZ Ceti stars. To this end,  we employed the {\tt LP-PUL} pulsation code to perform an exploratory pulsational investigation by computing adiabatic $g$-mode Newtonian pulsations on fully relativistic equilibrium white dwarf models. We 
find that general relativity impacts the Brunt V\"ais\"al\"a frequency profile of the most massive models, substantially altering their $g$-mode pulsational properties such as the forward period spacing and the kinetic energy of oscillation.

Summarizing, general relativity effects should be taken into account for an accurate assessment of the
structural, evolutionary, and pulsational properties of white dwarfs with masses above $\approx$ 1.30\ M$_{\sun}$. It should be appropriate to refer to such white dwarfs as {\it relativistic ultra-massive white dwarfs}. The relativistic CO ultra-massive white dwarf evolutionary sequences presented here constitute an improvement over those computed in the framework of Newtonian gravity.  The impact of general relativity on the thermo-mechanical structure of the most massive white dwarfs could affect the energetics of the explosion 
and the nucleosynthesis of neutron-rich matter (and thus be relevant for the better understanding of type-Ia supernova explosions) as well as the critical density that separates 
the explosive and the collapse outcomes of these supernovae.

\section*{Acknowledgements}
We thank Domingo Garc\'ia-Senz for valuable comments about the possible impact of our models on
  Type Ia Supernovae and
  Detlev Koester for providing atmosphere models to the high surface gravities that characterise our relativistic ultra-massive white dwarf models. 
Part of  this work was  supported by PICT-2017-0884 from ANPCyT, PIP
112-200801-00940 grant from CONICET, and by the Spanish project PID 2019-109363GB-100. MEC acknowledges Grant RYC2021-032721-I funded by MCIN/AEI/10.13039/501100011033 and by
the European Union NextGenerationEU/PRTR.
ST, RR and ARM acknowledge support from MINECO under the PID2020-117252GB-I00 grant and from the AGAUR/Generalitat de Catalunya grant SGR-386/2021. 
RR acknowledges support from Grant RYC2021-030837-I funded by MCIN/AEI/ 10.13039/501100011033 and by “European Union NextGenerationEU/PRTR”.
This  research has  made use of  NASA Astrophysics Data System. This work has made use of data from the European Space Agency (ESA) mission {\it Gaia} (\url{https://www.cosmos.esa.int/gaia}), processed by the {\it Gaia} Data Processing and Analysis Consortium (DPAC, \url{https://www.cosmos.esa.int/web/gaia/dpac/consortium}). Funding for the DPAC has been provided by national institutions, in particular the institutions participating in the {\it Gaia} Multilateral Agreement. 

\section*{Data Availability Statement}

Supplementary material will be available to all readers.
The cooling sequences are publicly available for download at
\href{ http://evolgroup.fcaglp.unlp.edu.ar/TRACKS/tracks.html}{http://evolgroup.fcaglp.unlp.edu.ar/TRACKS/tracks.html}.



\bibliographystyle{mnras}
\bibliography{ultramassiveCO.bib}


\bsp	
\label{lastpage}
\end{document}